\documentclass[conference]{IEEEtran}
\usepackage{cite}

\ifCLASSINFOpdf
\else
   \usepackage[dvips]{graphicx}
\fi
\usepackage[cmex10]{amsmath}
\usepackage{amsthm, amssymb}
\usepackage{array}

\usepackage{mdwmath}
\usepackage{mdwtab}
\newcommand{\RC}{$RC$}

\newcommand{\OLN}{$NC$ }
\newcommand{\MOLN}{$MAX-NC$ }

\newcommand{\KSAT}{$K-SAT$}

\newtheorem{theorem}{Theorem}
\begin{document}
%
\title{Maximizing the number of accepted flows in TDMA-based wireless ad hoc networks is APX-complete}


\author{\IEEEauthorblockN{Raffaele Bruno}
\IEEEauthorblockA{ITT-CNR\\
Via G. Moruzzi 1, 56124 Pisa - Italy\\
Email: \emph{r.bruno@iit.cnr.it}}
\and
\IEEEauthorblockN{Vania Conan and Stephane Rousseau}
\IEEEauthorblockA{Thales Communications\\
160 Boulevard de Valmy 
 ÐBP82 Ð92704 Colombes Cedex -France\\
Email: \emph{firstname.surname@fr.thalesgroup.com}}
}
\maketitle

\begin{abstract}
Full exploitation of the bandwidth resources of Wireless Networks is challenging because of the sharing of the radio medium among neighboring nodes. Practical algorithms and distributed schemes that tries to optimising the use of the network radio resources. In this technical report we present the proof that maximising the network capacity is is an APX-Complete problem (not approximable within $1/(1-2^{-k})-\epsilon$ for $\epsilon>0$).
\end{abstract}
\IEEEpeerreviewmaketitle

\section{Introduction}

Wireless networks are becoming attractive thanks to the ease  and  low cost of their installation. Moreover, they provide high capacity bandwdith and support client mobility. They tend to replace wired networks in urban areas. Thus, they have to support different multi-hop QoS and priority flows. 

Contrary to  wired networks, in wireless  network, when a node emits, all nodes in the emission  range  receive the signal transmission. When this emission disturbes other emissions we talk about interferences. This interference phenomena is hard to assess, some experimental studies are  done in~\cite{9,10,34}. Moreover, this phenomena is also hard to modelize~\cite{17,19,23,33}.
 
Due to interferences, some problems that can be solved in polynomial time  with a distributed algorithm in a wired networks  become NP-Complete in wireless networks. 

Herein, we investigate one of these problems that consists in maximizing the network capacity~\cite{35,4}. We give the theoretical proof that  this optimization problem is at least not approximable within $1/(1-2^{-k})-\epsilon$ for $\epsilon>0$ whatever the PHY/MAC layers used (either CSMA or TDMA). In orther words, we prove that this optimization problem is  APX-Complete. This main contribution concludes on the complexity of the QoS routing problem in wireless networks and  justifies the use of heuristics~\cite{6,7,8,11,12,14,28} for routing decisions.

The remainder of the paper is structured as follows: In Section~\ref{sec: th}, we give both network and interferences models, we define the  optimization problem. Then, we present the proof that this optimization problem is APX-complete for particular instances at least not approximable within $1/(1-2^{-k})-\epsilon$ for $\epsilon>0$.  Finally, in Section~\ref{sec: con}, we conclude the paper and provide outlines our on-going work.
\subsection{Related Work}
To achieve the maximimum throughput capacity of a network several optimization problems have to be solved. In~\cite{43}, they investigate the problem of channel assignement in wireless network. They prove that this problem is NP-Complete and provide a polynomial-time approximation scheme (abbreviated PTAS) for this problem. The network capacity depends also on the number of time slots required to successfully schedule all links. In~\cite{42}, authors investigate the scheduling problem with SINR constraints, based on the physical SINR models -also called PHY graph model-~\cite{37,38}, and show it to be NP-Complete. In order to prove this result, they give a polynomial time reduction from  the well-known subset sum problem. The approximation corresponding problem, called Approx-Subset-Sum is an FPTAS. In~\cite{41}, authors also invesgate the throughput maximization problem under SINR constraints model and graph-model. They conjecture that the throughput maximization problem is NP-Complete according to the result given in~\cite{42}. In this article, we prove the validity of this conjecture and moreover, we provide an additional result that is the approximation throughput maximization problem is APX-Complete. More precisly, we focus on the Remaining Capacity problem (\RC) as defined in~\cite{4}.

In a wired  network, finding an elementary path between two nodes minimizing over-loaded nodes can be solved by using the distributed and polynomial dijkstra algorithm~\cite{5}. In a wireless  network, this problem is the Remaining Capacity problem (\RC) defined in~\cite{4}. In~\cite{1,2}, authors show 
that finding the shortest path (repectively longest path) that avoid over-loaded nodes cannot be solved in a polynomial time. Centralized heuristics are proposed in~\cite{6}. Moreover, experimental studies aim at increasing the capacity of the network~\cite{13} (i.e. we call network capacity the cumulated data rate flows present in the network)  by giving the best routing decisions (distributed routing protocols) that avoids using over-loaded nodes, see ~\cite{6,7,8,11,12,14}. In~\cite{28}, authors try to decreased the maximum load  with curve routing. This routing needs the knowledge of the geographical localization of nodes. 

Herein, we prove that whatever  PHY/MAC (e.g. CSMA, TDMA) assumptions and whatever interference model (e.g. graph-based model, PHY graph model) given, in a single radio network, the problem of maximizing the network capacity  is APX-Complete. 

\section{Theoretical Study}\label{sec: th}
In this section, we precise the network and interferences models we consider, see Section~\ref{subsec: th:model}. Then, we give the definition of the optimization problem, see Section~\ref{subsec: th:definition}. In Section~\ref{subsec: th:proof}, we give the proof that this problem is APX-Complete.  
\subsection{Network and interferences models}
\label{subsec: th:model}
Let us consider the single radio wireless network $\mathcal{N}$ as a undirected graph $G(V,E,w)$. For each nodes $N_i$ in the wireless network $\mathcal{N}$ corresponds a node (or vertex) $n_i$ in $V$. Moreover, in the wireless network $\mathcal{N}$, if the node $N_i$ is in the transmission range of the node $N_j$ then $n_i$ and $n_j$ are linked by an edge in the $G$.

Each node can emit or receive a fixed number of bandwidth units (called capacity of the node) given by the weighted function $w:V\rightarrow \mathbb{N}$. At each step (time-slot) each node can emit (or receive) one packet unit to (or from) one node in its neighborhood. Moreover when a node transmits data flow to another one, all nodes in its neighborhood receive this data flow. Then, their remaining capacity decreases. When the remaining capacity of a node is equal to zero then the node cannot emit or receive data flow any more. If the remaining capacity is lower than zero, this node is over-loaded.
\begin{figure}[!h]
\begin{center}
\includegraphics[width=3.25in]{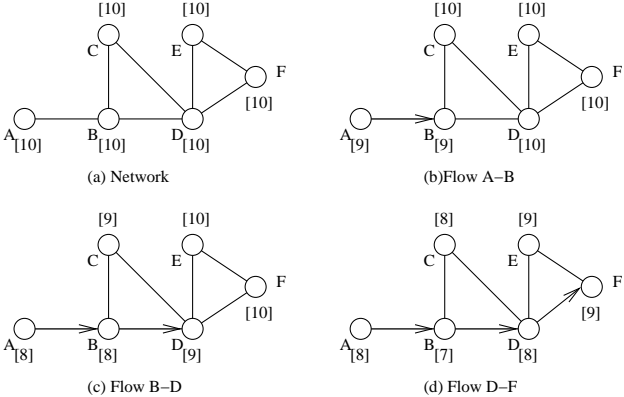}
\caption{\label{FIG : exampleNetwork} Example of a flow transmission}
\end{center}
\end{figure}
In Figure~\ref{FIG : exampleNetwork}.a, we represent a graph $G(V,E,w)$ composed of $6$ nodes ($\{A,\ B,\ C,\ D,\ E,\ F\}$). The initial capacity (in square brackets) is equal to $10$ for all nodes. Thus we have $\forall n\in V,\ w(n)=10$. A flow has to be routed from $A$ to $F$. In Figure~\ref{FIG : exampleNetwork}.b, the node $A$ transmits the data flow to $B$. Their respective remaining capacity is decreased by $1$. In Figure~\ref{FIG : exampleNetwork}.c, the node $B$ transmits the data flow to $D$ while the node $A$ still transmits the data flow to $B$. Then the remaining capacity of $A$, $B$, $C$ and $D$ is decreased by $1$. Finally, in Figure~\ref{FIG : exampleNetwork}.d the node $D$ transmits the data flow to $F$ while the two other transmissions carry on. The capacity of nodes $B$, $C$, $D$,$E$ and $F$ is decreased by $1$.

In this model, we assume that links are bi-directionnal, symetrical and the quality of transmission is maximal. Nodes in this slotted network are synchronized.
this model is very favorable inasmuch as  accurate timing synchronization within a global network is difficult to achieve~\cite{25}. In the next section, we prove that even in this favorable context, the problem maximizing the network capacity  is APX-Complete. This problem becomes certainly more difficult under real conditions.

\subsection{Problems definition}
\label{subsec: th:definition}
To prove that it is also not approximable, we will work on a subset of the problem $MAX \!\!-\!\! NC$, or \emph{instance} of the problem, defined as:
\begin{itemize}
\setlength{\itemsep}{6pt}
\item{\bf Instance}: An undirected graph $G=\left(V,E,L,w\right)$, where $V$ is a set of nodes and $E$ is a set of edges. On each node $k \!\in\! V$ a TDMA-Frame of $L(k)$ slots with a duration equal to $\tau=1$. To each edge $e\in E$ is associated a weight $w(e)=1$. An infinite set $\mathcal{F}$ flows $f_1,...,f_i, f_{i+1},.., f_\infty$  from $s_i$ to $d_i$. All flows require a bandwidth equivalent to $1$ slot.
\item{\bf Solution}: An elementary path set $\mathcal{P}_n$ for a subset of flows $F_n={f_1,\ ...,\ f_i,\ f_{i+1},\ f_n}\subset \mathcal{F}$ such that there exists a path $p_i$ for each flow $f_i\in F_n$ that connects $s_i$ and $d_i$ and along which all slots can be assigned.
\item{\bf Measure}: Value of accepted flows $n$.
\end{itemize}

\label{sec:forth}
\begin{theorem}\label{theo : proof KSAT}
The problem of Maximizing Network Capacity
in a TDMA-based ad hoc network is APX-complete, i.e.,
it is not approximable because there is no polynomial-time
approximation scheme.
\end{theorem}
\label{subsec: th:proof}

\begin{proof}
First, it is trivial that the decision problem \OLN ~~associated to the optimization problem \MOLN ~~is NP-Complete by extension of the NP-Complete Remaining Capacity problem, see~\cite{4}. In order to prove the \MOLN ~~optimization problem to be APX-Complete we first reduce K-Satisfiability  decision problem (\KSAT) to the associated  \OLN ~~decision problem by transforming any instance $\mathcal{I}=(U,C)$ of the \KSAT ~~problem to an instance $\mathcal{I'}=(G,\mathcal{F}, F)$ of \OLN ~~problem. This means we demonstrate how to convert clauses that contain from $3$ to $k$ literals (boolean variables) into a particular instance of \OLN ~~problem. \\

In the second part of the proof, we conclude on the  APX-Completeness by proving that maximizing the number of clauses satisfied by the truth assignment in the MAX \KSAT ~~problem means maximizing the network capacity in the \MOLN ~~problem.\\

Here is the definition of the {\bf MAX K-Satisfiability  problem} (MAX \KSAT):
\begin{description}
\item {\bf K-Satisfiability  problem}:
\item \begin{itemize}
\item{\bf Instance}: Set $U=\{u_1,\ ...,\ u_n\}$ of $n$ variables, collection $C=\{c_1,\ ...,\ c_m\}$ of $m$ disjunctive clauses of at most $k$ literals, where a literal is a variable or a negated variable in $U$. k is a constant, $k\ge 2$.({\it n.b. we note $\mathcal{I}=(U,C)$ an instance of the \OLN ~~problem.}).
\item{\bf Solution}: A truth assignment for $U$.
\item{\bf Measure}: Number of clauses satisfied by the truth assignent.
\end{itemize}\end{description}

First, from the $m$ clauses $c_i\in C$ of the MAX \KSAT ~~problem, we build a graph $G(V,E,w)$ of the instance $\mathcal{I'}$ of \OLN ~~problem. Nodes of $V$ can be divided into five subsets:
\begin{itemize}
\item {\bf subset~1 (V1)} consists of $2\times m$ nodes:
\begin{equation}\label{eqOLN : V1}
\begin{array}{lcl}
V1&=&\{n_1^1,n_4^1,n_1^2,n_4^2,...,n_1^n,n_4^n\}\\
\end{array}
\end{equation}
\item {\bf subset~2 (V2)} consists of $\sum_{i=1}^{i\leq m}|c_i|$ nodes:
\begin{equation}\label{eqOLN : V2}
\begin{array}{lclll}
V2&=&\{n_6^1,\ n_9^1,\ n_{4+3\times|c_1|-1}^1,\\
&&...,\\
&&\ n_6^{i},\ n_9^i,\ n_{4+3\times|c_i|-1}^i,\\
&&...,\\
&&\ n_6^{m},\ n_9^m,\ n_{4+3\times|c_n|-1}^m\}\\
\end{array}
\end{equation}
For each literal of each  clause $c_i$ corresponds one and only one node in {\bf V2}. 
\item {\bf subset~3 (V3)} consists of $2\times\sum_{i=1}^{i\leq m}|c_i|$ nodes:
\begin{equation}\label{eqOLN : V3}
\begin{array}{lclll}
V3&=&\{n_5^1,\ n_8^1,\ ...,\ n_{4+3\times|c_1|-2}^1,\\
&&n_7^1,\ n_{10}^1,\ ...,\ n_{4+3\times|c_1|}^1,\\
&&...\\
&&n_5^i,\ n_8^i,,\ ...\ n_{4+3\times|c_i|-2}^i,\\
&&n_7^i,\ n_{10}^i,,\ ...\ n_{4+3\times|c_i|}^i,\\
&&...,\\
&&n_5^n,\ n_8^m,,\ ...\ n_{4+3\times|c_m|-2}^m,\\
&&n_7^n,\ n_{10}^m,,\ ...\ n_{4+3\times|c_m|}^m\}\\
\end{array}
\end{equation}
\item {\bf subset~4 (V4)} consists of $m$ nodes:
\begin{equation}\label{eqOLN : V4}
\begin{array}{lclll}
V4&=&\{n_{4+3\times(|c_1|+1)-2}^1,\\
&& ...\\
&&n_{4+3\times(|c_i|+1)-2}^i,,\\
&&...\\
&&n_{4+3\times(|c_n|+1)-2}^m\}\\
\end{array}
\end{equation}
\item {\bf subset~5 (V5)} \\consists of $\sum_{i=1}^{n}(\sum_{j=1}^{m} {\bf 1}_{u_i^j} \times \sum_{j=1}^{m} {\bf 1}_{\neg u_i^j})$ nodes:
\begin{equation}\label{eqOLN : V5}
\begin{array}{lclll}
V5&=&\{n_{1},n_2,...,n_{\sum_{i=1}^{n}(\sum_{j=1}^{m} {\bf 1}_{u_i^j} \times \sum_{j=1}^{m} {\bf 1}_{\neg u_i^j})}\}\\
\end{array}
\end{equation}
\end{itemize}
To illustrate the contruction of the set of nodes $V=\{\cup_{i=1}^{i\leq 5}Vi\}$ of the graph $G=(V,E,w)$, we propose an example by reducing the following formula~\ref{eqOLN: example}, according to the eq~\ref{eqOLN : V1}, eq~\ref{eqOLN : V2}, eq~\ref{eqOLN : V3}, eq~\ref{eqOLN : V4} and eq~\ref{eqOLN : V5}. 

\begin{equation}\label{eqOLN: example}
(a \wedge b \wedge c \wedge d)\vee(\neg a \wedge b \wedge e \wedge f)\vee(\neg a \wedge \neg b \wedge \neg c \wedge \neg d)
\end{equation}
In Figure~\ref{FIG : example20a}, we give the set of nodes $V$ of the instance of the \OLN ~~decision problem from the forumla~\ref{eqOLN: example}. \\

\begin{figure}[!h]
\begin{center}
\includegraphics[width=3.25in]{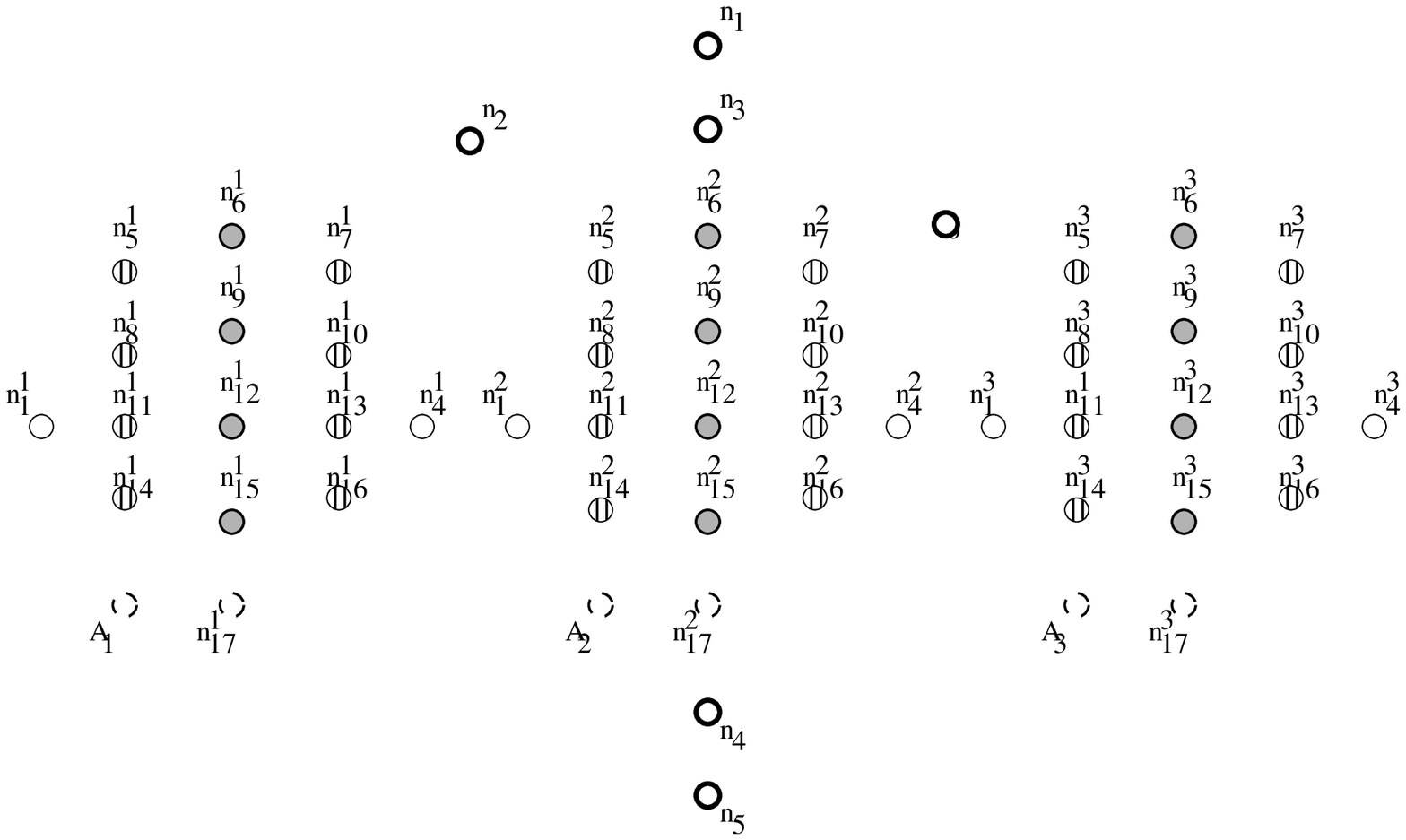}
\caption{\label{FIG : example20a} Nodes of $G(V,E,w)$}
\end{center}
\end{figure}

We distinguish the different sets of nodes :
\begin{description}
\item \includegraphics[width=0.2cm]{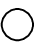} represents the set of nodes $V1$
\item \includegraphics[width=0.2cm]{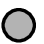} represents the set of nodes $V2$
\item \includegraphics[width=0.2cm]{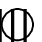} represents the set of nodes $V3$
\item \includegraphics[width=0.2cm]{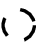} represents the set of nodes $V4$
\item \includegraphics[width=0.2cm]{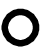} represents the set of nodes $V5$
\end{description}

~~\\
We continue the construction by adding the edges. We define the set $E$ of edges of $G(V,E,w)$:
\begin{itemize}
\item {\bf subset~1 (E1)} :
\begin{equation}\label{eqOLN : E1}
\begin{array}{lclll}
E1&=&\{\cup_{i=1}^{i\leq m}(\cup_{j=1}^{j\leq |c_i|}(n_1^i,n_{4+3\times j -2}^i)),\\
&&\cup_{i=1}^{i\leq m}(\cup_{j=1}^{j\leq |c_i|}(n_{4+3\times j -2}^i,n_{4+3\times j-1}^i)),\\
&&\cup_{i=1}^{i\leq m}(\cup_{j=1}^{j\leq |c_i|}(n_{4+3\times j-1}^i,n_{4+3\times j}^i)),\\
&&\cup_{i=1}^{i\leq m}(\cup_{j=1}^{j\leq |c_i|}(n_{4+3\times j}^i,n_4^i)),\\
&&\cup_{i=1}^{i\leq m}(n_1^i,n_{4+3\times (|c_i|+1)-2}^i),\\
&&\cup_{i=1}^{i\leq m}(n_{4+3\times (|c_i|+1)-2}^i,n_4^i),\\
\end{array}
\end{equation}
\item  {\bf subset~2 (E2)} :
\end{itemize}
\begin{equation}\label{eqOLN : E2}
\begin{array}{llll}
E2=&\{\cup_{i=1}^{i\leq m}(\cup_{j=1}^{j\leq |c_i|-1}(\cup_{k=j+1}^{k\leq |c_i|}(n_{4+3\times j -1}^i,n_{4+3\times (k) -1}^i))),\\
&\cup_{i=1}^{i\leq m}(n_{4+3\times (|c_i|+2)-2}^i,n_{4+3\times (|c_i|+2)-1}^i\}
\end{array}
\end{equation}
\begin{itemize}
\item {\bf subset~3 (E3)} :
\begin{equation}\label{eqOLN : E3}
\begin{array}{lclll}
E5&=&\{\cup_{i=1}^{i\leq m-1}(n_4^i,n_1^{i+1})\}
\end{array}
\end{equation} 
\item {\bf subset~4 (E4)} : consists of linking each node $n_i\in S6$ with a pair of nodes $(n_{k1}^i,n_{k2}^{j\neq i})$ such as either the literal associated to the node $n_{k1}^i$ is on the non-complementary form and the one associated to $n_{k2}^{j\neq i}$ is on the complementary form, or $n_{k1}^i$ is on the complementary form and the one associated to $n_{k2}^{j\neq i}$ is on the non-complementary form.
\end{itemize}

Figure~\ref{FIG : example20j} represents the set of edges $E$ of the graph $G(V,E,w)$. We distinguish the four sets of edges :
\begin{description}
\item \includegraphics[width=1cm]{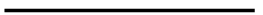} represents the set of edges $E1$
\item \includegraphics[width=1cm]{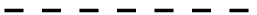} represents the set of edges $E2$
\item \includegraphics[width=1cm]{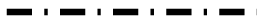} represents the set of edges $E3$
\item \includegraphics[width=1cm]{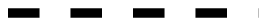} represents the set of edges $E4$
\end{description}

\begin{figure}[!h]
\begin{center}
\includegraphics[width=3.25in]{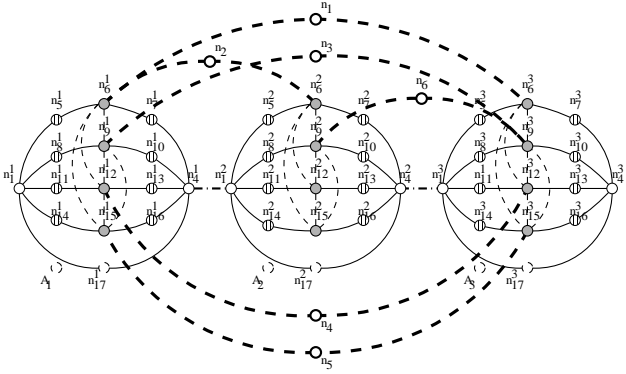}
\caption{\label{FIG : example20j} Edges of $G(V,E,w)$}
\end{center}
\end{figure}

We now define the weight function $w:V\rightarrow \mathbb{N}$.
\begin{equation}\label{eqOLN : W1}
w\left \{
\begin{array}{c c @{=} l}
\forall v\in V1\cup V3,&\ w(v)\ &3\\
\forall v\in V2,&\ w(v)&5,\\
\forall v\in V4&\ w(v)&3,\\
\forall v\in V5,&\ w(v)\ &1
\end{array}
\right.
\end{equation}

Figure~\ref{FIG : example20f} represents the capacity of each node -given by Equation~\ref{eqOLN : W1}- in square barkets.
\begin{figure}[!h]
\begin{center}
\includegraphics[width=3.25in]{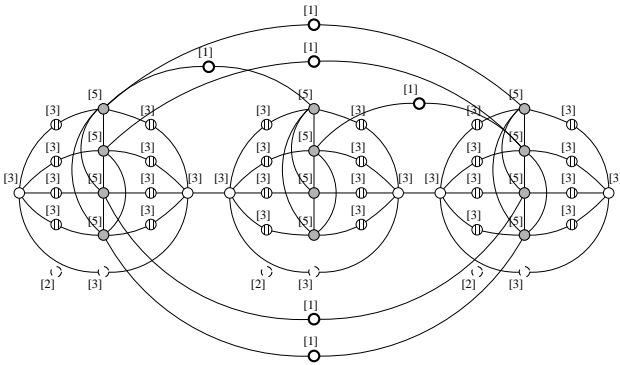}
\caption{\label{FIG : example20f} Capacity of each nodes}
\end{center}
\end{figure}
Let $\mathcal{F}$ be an infinite set of flows such as $f_1$ has to connect $A_1$ and $n^1_{17}$, $f_2$ has to connect $A_2$ and $n^2_{17}$, $f_3$ has to connect $A_3$ and $n^3_{17}$ and all flows $f_{3\leq i \leq \infty}$ have to connect $n^1_1$ and $n^3_4$. 

Maximizing the number of accepted flows consists in first finding a path for flow $f_1$, then, for flow $f2$ and so on. It is trivial to note that the first three flows $f_1$ $f_2$ and $f_3$ can be routed. The problem appears for the forth flow $f_4$, the one from $n^1_1$ to $s^3_4$. The key idea behind the proof is the following: if this flow $f_4$ can be routed without pre-empting any of the two first flows then a true assignment exists for the $MAX \; 3 \!\!-\!\! SAT$ problem. If no path can be found to route the flow $f_3$ without pre-empting either $f_1$ or $f_2$ or $f_3$, then no true assignment for the $MAX \; 3 \!\!-\!\! SAT$ problem can be found. These two statements would ensure a one-to-one mapping of both instances of the problem.

Thus, we give the construction of an instance $\mathcal{I'}$ of the \OLN ~~problem from an arbitrary instance $\mathcal{I}$ of the \KSAT ~~problem. The three first flows can be easily routed (only one route exists).
The \OLN~~problem consists in finding a path $P$ for a forth  flow $F$ of data rate equal to $1$ unit from the node $n_1^1$ to the node $n_4^m$ respecting the constraint given in Equation~\ref{eq: OLN2}. It is possible to route this new  flow accross the path :
\begin{equation*}
\begin{array}{ll}
P=\{&n_1^1,\ n_{4+3\times (|c_1|+2)-1}^1,\ n_4^1,\ n_{4+3\times (|c_i|+2)-1}^i,\ ...,\\&n_1^m,\ n_{4+3\times (|c_m|+2)-1}^m,\ n_4^m\}
\end{array}
\end{equation*}
by over-loading all nodes of the subset $V4$. However, this is not a solution for the \OLN ~~problem in which none of nodes can be over-loaded. Then to reach  the node $n_4^m$ from the node $n_1^1$, the new flow has to be routed via nodes from the subset $V2$ and not from the subset $V4$.\\

Let $\mathcal{S}_{K-SAT}$ be a solution of the \KSAT ~~problem. We detail how we  obtain a corresponding solution $\mathcal{S}_{NC}$ for the \OLN ~~problem. For each literal that is "true" for the solution $\mathcal{S}_{K-SAT}$, we add the corresponding node of $V2$ in the path solution $\mathcal{P}_{NC}$ of the \OLN ~~problem:
 We add also all nodes of the set $V1$. We complete the path by adding nodes $n_{k}^i\in V3$ and $n_{k'}^i\in V3$ such as:
$$k=(\min_{i=1}^{|c_i|}n_{4*3\times i -1} \in \mathcal{P}) - 1$$
$$k=(\max_{i=1}^{|c_i|}n_{4*3\times i -1} \in \mathcal{P}) + 1$$
Here is a true assignment $\mathcal{A}_1$(a solution) in the example~\ref{eqOLN: example} for the \KSAT~~problem and a wrong assignment $\mathcal{A}_2$:
\begin{equation}\label{eqOLN: solution KSAT S1}
\begin{array}{lcl}
\mathcal{A}_1&=\{&a=true;b=true;c=true;
\\&&d=false;e=false;f=false\}
\end{array}
\end{equation}

\begin{equation}\label{eqOLN: non-solution KSAT S2}
\begin{array}{lcl}
\mathcal{A}_2&=\{&a=true;b=true;c=true;\\&&d=true;e=false;f=false\}
\end{array}
\end{equation}

Here are the corresponding paths $\mathcal{P}_1$ and $\mathcal{P}_2$ for the \OLN ~~problem:
\begin{equation}\label{eqOLN: solution WQP S1}
\begin{array}{ll}
\mathcal{P}_1=\{&n_1^1,n_5^1,n_6^1,n_9^1,n_{12}^1,n_{13}^1,n_4^1,n_1^2,n_8^2,n_9^2,n_{10}^2,n_{4}^2,\\&n_1^3,n_{14}^3,n_{15}^3,n_9^3,n_{16}^3,n_4^3\}
\end{array}
\end{equation}
\begin{figure}[!h]
\begin{center}
\includegraphics[width=3.25in]{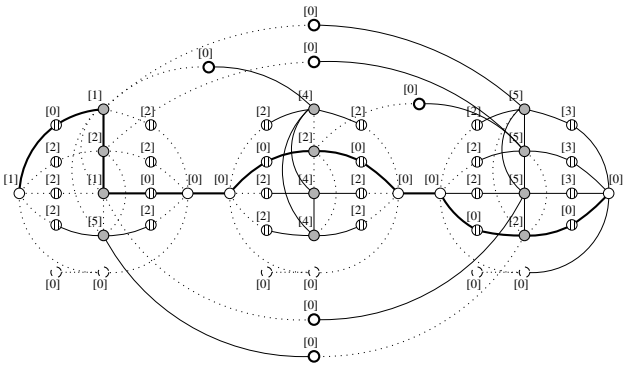}
\caption{\label{FIG: example20h.eps} Path that corresponds to the assignment given in equation~\ref{eqOLN: solution WQP S1}}
\end{center}
\end{figure}

\begin{equation}\label{eqOLN: non-solution WQP S2}
\begin{array}{ll}
\mathcal{P}_2=\{&n_1^1,n_5^1,n_6^1,n_9^1,n_{12}^1,n_{15}^1,n_{16}^1,n_4^1,\\&n_1^2,n_8^2,n_9^2,n_{10}^2,n_{4}^2,n_{8}^3,n_4^3\}
\end{array}
\end{equation}
\begin{figure}[!h]
\begin{center}
\includegraphics[width=3.25in]{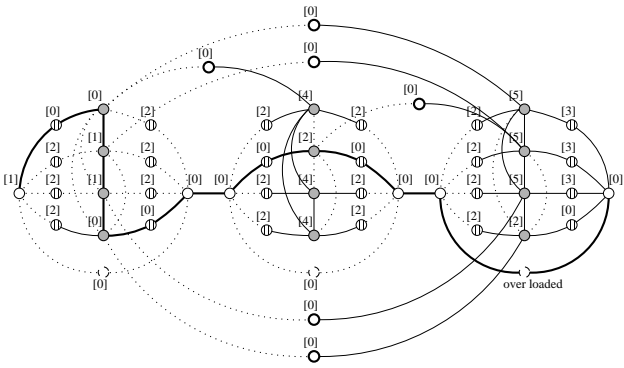}
\caption{\label{FIG: example20i.eps} Path that corresponds to the assignment given in equation~\ref{eqOLN: non-solution WQP S2}}
\end{center}
\end{figure}
In $\mathcal{P}_1$ none of the flows from $\mathcal{F}$ are preemted whereas is $\mathcal{P}_2$ one of them ($f_3$) is preemtped. Thus, $\mathcal{P}_1$ is a solution for the \OLN ~~problem and $\mathcal{P}_2$ is not.\\

We prove now that a solution for the \KSAT ~~problem is a solution for the \OLN ~~problem and, conversely, that a non-solution for the \KSAT~~problem is not a solution for the \OLN ~~problem.
\begin{itemize}
\item {\bf A solution for the \KSAT ~~problem is a solution for the \OLN ~~problem: } A solution for the \KSAT ~~problem is an assignment of each variable such as the formula is true. The only constraint in this assignment is that when a literal is  "true" in one of the clause, then this literal  is true for all clauses where it appears in the same form (complementary or not) and false otherwize. According to the construction, when a literal is "true" in the solution of the \KSAT ~~problem, then the corresponding node in the graph $G(V,E,w)$ belongs to the solution path of the \OLN ~~problem. None of the nodes from the set $V5$ can be saturated as far as two literals cannot belong to the solution of the \KSAT ~~problem if the corresponding variable is not in the form (complementary/non-complementary).
\item {\bf If the assignment $\mathcal{A}$ is not a solution for \KSAT ~~problem, then the associated path is not a solution for the \OLN ~~problem: }an assignment such as the formula is not true implies that at least one clause $c_i$ of the $m$ clauses is not true. Then, it means that in the associated path $\mathcal{P}$  it is not possible to find a path from $n_1^i$ to $n_4^i$ respecting the constraint given in Equation~\ref{eq: OLN2} and without over-loading any nodes. Indeed, if all literals are false, it means that the new flow cannot be routed via any nodes from $V2$ without over-loading any nodes.

\end{itemize}

The second part of the proof consists in showing that maximizing the number of clauses satisfied by the truth assignment in the \KSAT ~~problem means maximizing the network capacity in the \OLN ~~problem. \\

When the clause $c_i$ cannot be satisfied in the \KSAT ~~problem, then no path can be found between $n_1^i$ and $n_4^i$ in the \OLN ~~problem. Then, the only way to connect these two nodes is to over-load at least one node (e.g. the node $n_{4+3\times(|c_i|+1)-2}$). When the clause $c_i$ can be satisfied, it means that there exists a path from $n_1^i$ and $n_4^i$. Thus, maximizing the number of satisfied clauses means maximizing the network capacity. 

Recall that the MAX \KSAT ~~problem is not approximable within $1/(1-2^{-k}-\epsilon)$ for any $\epsilon > 0$ and $k\leq 3$. Then, we find a particular instance of the \OLN ~~problem for which the difficulty is at least the same as the MAX \KSAT ~~problem where $k\leq 3$. We can conclude that we can find instances for which  the \MOLN ~~problem is not approximable  within $1/(1-2^{-k}-\epsilon)$ for any $\epsilon > 0$.

This concludes the proof.
 
\end{proof}

\section{conclusion}\label{sec: con}
We prove that maximizing the network capacity is  APX-Complete by reduction of the \KSAT~~problem. Moreover, this proof is given for a very favorable context, and this problem becomes more difficult if we consider the real constraints (distributed solution for scalability, synchronization of nodes, ...). Thus this result is a very important one that justifies the use of heuristic in routing decisions.

\section*{Acknowledgment}
This work was partially funded by the European Commission Programme ICT-2008-215320 through the EU-MESH Project.

\end{document}